\documentstyle[aps,prl,twocolumn,epsfig]{revtex}
\newcommand{\BE}{\begin{equation}}
\newcommand{\EE}{\end{equation}}
\newcommand{\bx}{{\bf x}}

\newcommand{\bv}{{\bf v}}
\newcommand{\etab}{\mbox{\boldmath $\eta$}}

\begin{document}
\draft
\title{Diffusive transport and self-consistent dynamics in
  coupled maps}
%\title{Transport in active systems: diffusion in self-consistent coupled maps}
%\title{Diffusion in self-consistent systems:
%mean field coupling of symplectic maps}
%\title{Diffusive properties of self-consistent systems described
%by symplectic maps}
\author{Guido Boffetta$^1$, Diego del-Castillo-Negrete$^2$,
Crist\'obal L\'opez$^3$, Giuseppe Pucacco$^4$ and Angelo Vulpiani$^5$
}
\address{$^1$ Dipartimento di Fisica Generale and INFM,
Universit\`a di Torino, \\
via P. Giuria 1, 10125 Torino, Italy}
\address{$^2$ Oak Ridge National Laboratory,
Oak Ridge TN, 37831-8071, USA}
\address{$^3$ Dipartimento di Fisica,
Universit\`a di Roma ``La Sapienza'', \\
p.le A. Moro 2, 00185 Roma, Italy}
\address{$^4$ Dipartimento di Fisica,
Universit\`a di Roma ``Tor Vergata'', \\
via della Ricerca Scientifica 1, 00133 Roma, Italy}
\address{$^5$ Dipartimento di Fisica and INFM  UdR and SMC Center,
Universit\`a di Roma ``La Sapienza'', \\
p.le A. Moro 2, 00185 Roma, Italy}
\date{\today}
\maketitle

\begin{abstract}
The study of diffusion in Hamiltonian systems has been a problem of
interest  for a number of years.
   In this paper  we explore the influence of  self-consistency on
the diffusion properties of systems described by coupled symplectic maps.
Self-consistency, i.e. the back-influence of the transported quantity on the
velocity field of the driving flow,  despite of its critical
importance,  is usually overlooked in the description of
realistic systems, for example in plasma physics.
We propose a class of  self-consistent models consisting of an ensemble of
maps  globally coupled through a mean field.
Depending on the kind of coupling, two different general types of
self-consistent maps are considered: maps coupled to the field only
through the phase, and fully coupled maps, i.e. through the phase and
the amplitude of the external field.
The analogies and differences of the diffusion properties of these two kinds of
maps are discussed in detail.
\end{abstract}


%\begin{multicols}{2}

%%%%%%%%%%%%%%%%%%%%%%%%%%%%%%%%%%%%%%%%%%%%%%%%%%%%%%%%%%%%%%%%%%%%%%%
\section{Introduction}
\label{sec:1}

Understanding transport  is a problem of considerable
practical and theoretical interest in a great variety of fields ranging
from geophysics to chemical engineering and plasma physics.
In some situations one can safely consider the simple case
of {\em passive transport}  in which the transported quantity
does not affect the advecting flow~\cite{moffat,ottino}.
In the case of a scalar passive
field $\Theta({\bf x},t) $, the evolution equation is the
advection-diffusion equation
\BE
{\partial \Theta  \over \partial t}+\nabla \cdot (\bv \Theta)
=D_0 \nabla^2 \Theta,
\label{eq:1.1}
\EE
where ${\bf v}({\bf x},t)$ is the  velocity field, $D_0$
the molecular diffusivity, and $\Theta$ represents the scalar
concentration,  e.g. the
temperature of the fluid or  the concentration of a pollutant.
The domain of applicability of Eq.~(\ref{eq:1.1}) is limited by two
important physical
assumptions:
$\Theta$ has to be {\it inert}  (no possible chemical or biological
reactions are considered) and passive (there is no feedback on
the velocity). Reactive processes can be taken into account  by adding
to the right hand side  of Eq.~(\ref{eq:1.1})
a function $f(\Theta)$ modeling the reaction kinetics~\cite{xin}. This leads to
the so-called
advection-reaction-diffusion (ARD) equations widely used in the modeling
of chemical and biological systems including combustion,  diluted
chemical reactions, and
population dynamics~\cite{clopez}.

Taking into account the feedback of $\Theta$ on $\bv$, i.e., the
problem of {\em active
transport},  is in general more complicated as this involves
the equation of motion
for $\bv$.  Because of this, active transport is also called {\em
self-consistent
transport}. A well-known example is  fluid convection in  the
Boussinesq approximation~\cite{lamb}.  In this case, $\Theta$ represents the
fluid temperature which
is an active scalar in the sense that it modifies the velocity field
through the buoyancy
force in the Navier-Stokes equation for ${\bf v}$.

The goal of the present paper is to study the problem of diffusion of
active scalars.
In particular, we are interested in the relationship between
self-consistent chaos and diffusion due to chaotic advection.
The study of diffusion requires an accurate numerical integration
of the equations of motion for very long times and many initial
conditions. A common strategy to bypass this technical
difficulty is to describe the time-continuous equations of motion
with  a discrete-time map. Here we follow this approach and study the
problem of diffusion in self-consistent symplectic maps.
In the remaining  of this introduction we discuss in some
detail  the problem of self-consistent transport in  fluids and
plasmas. The intent of this discussion is to provide a physical
motivation for the use of  globally coupled
maps for studying self-consistent transport.

One of the simplest physical examples of active transport is
two-dimensional incompressible flows. This motion is described by
Navier--Stokes equation (\ref{eq:1.1}) in which $\Theta$ represents
the vorticity, $\zeta={\bf \nabla}\times {\bf v}$.
Plasma physics is another area in which the problem of
self-consistent transport is crucial. For example, in the Vlasov
description of an electron plasma  \cite{nicholson} (in a uniform
neutralizing ion background) the system is described by the phase space
electron distribution function
$f$ which, for a one-dimensional system,  evolves according to the
Liouville equation
\begin{equation}
\label{eq:1.4}
\partial_t f+ u\, \partial_x f + \partial_x \phi \, \partial_u f=D_0
\partial^2_u f\, ,
\end{equation}
where the term on the right hand side is a Fokker-Plank collision operator, and
$(x,u)$ are the phase space coordinates.  This equation is analogous
to Eq.~(\ref{eq:1.1}) if one identifies $\Theta$ with $f$, and ${\bf v}$
with the transport velocity in phase space
$(u, \partial_x\, \phi)$. In this case the self-consistent coupling
is provided by the Poisson equation
\begin{equation}
%\frac{d^2 \phi}{d x^2} = Q(x,t)\, ,
\label{eq:poisson}
\partial^2_x\, \phi = \int f(x,u,t)\, d u - 1\, ,
\end{equation}
where the right hand side  is the charge distribution including the fixed
neutralizing ion background.
That is, the dynamics of an electron plasma is an active transport
problem in which the
transport velocity of the distribution function $f$ in phase space is
determined by $f$
through Poisson's equation.

The previous ideas on  self-consistent transport can be reformulated within the
Lagrangian description according to which transport is described in terms of
individual particle trajectories instead of scalar fields and
distribution functions. The Lagrangian description is important because it
is the natural description to formulate the
self-consistent transport problem in terms of symplectic maps which
are the main objects of study in the present paper.
As it is well-known~\cite{chandra}, the Lagrangian formulation of
(\ref{eq:1.1}) is the Langevin equation
\BE
\frac{d\bx}{dt}=\bv (\bx,t)+\sqrt{2D_0}\etab (t),
\label{eq:1.6}
\EE
describing
the motion of a test particle (the tracer),
where $\etab$ is a normalized Gaussian white noise with zero mean and
delta correlated in time:
\BE
<\eta_i(t)\eta_j(t')>=\delta_{ij}\delta(t-t').
\label{eq:1.7}
\EE
The passive scalar nature of $\Theta$ in Eq.~(\ref{eq:1.1}) reflects
in the absence of
coupling  in the Lagrangian equations of motion (\ref{eq:1.6}).  However,
in an active transport
problem, the self-consistent coupling between the field and the
transport velocity leads
to a nonlinear coupling between the Lagrangian equations of motion.
The fact that particles interact (and usually through long-range interaction)
implies that the phase space evolution  of particle $n$, depends on
the position
of all the $N$ particles
\BE
{d {\bf x}_{n}(t) \over d t} = {\bf v}({\bf x}_{1}(t), ..., {\bf x}_{N}(t)),
\label{eq:1.8}
\EE
and therefore the system has a phase space of dimension proportional to $N$.
This is the well-known $N$-body problem that arises in many fields of
physics, including gravitational systems in Astronomy~\cite{Giuseppe},
point vortices in two-dimensional fluid dynamics~\cite{aref1}, and
atomic physics.

An approximation of the $N$-body problem (\ref{eq:1.8}), which
is often used, is a mean field type approximation in which the
interaction among particles occurs through a global variable
${\bf X}$ function of all the particles.
  In the examples shown below,
the mean field will depend on the mean distribution of particles
only, thus (\ref{eq:1.8}) formally reduces to
\begin{eqnarray}
\label{eq:1.10}
{d {\bf x}_{n} \over d t} &=& {\bf v}_{ext}({\bf x}_n) +
{\bf v}({\bf x}_{n}-{\bf X}), \nonumber \\
& & \label{eq:1.5} \\
{d {\bf X} \over d t} &=& {\bf F}\left({\bf X}, \{ {\bf x}_k \} \right),
\nonumber
\end{eqnarray}
where we have included the possible contribution of an
external field ${\bf v}_{ext}$.

Recently, a mean-field description of this sort has been proposed to
study self-consistent transport in fluids and
plasmas~\cite{diego_CHAOS_00,diego_phys_a_00,diego_CHAOS_02}.
In the plasma physics context, this
approximation known
as the  single-wave-model (SWM), consists of simplifying the
self-consistent coupling between $f$ and $\phi$ given by
Poisson's equation (\ref{eq:poisson}).

The SWM is a  general model for the description of
marginally stable fluids and plasmas~\cite{diego_pop_98,neil_01b}. 
Also the model bears
many  interesting analogies with coupled oscillator models used in statistical
mechanics~\cite{antoni-ruffo-1995}. As  such,  it is an
insightful model to explore the problem of self-consistent chaos, and
will be our starting point for the construction of the 
self-consistent  symplectic map
models in the present paper.

The transition to the symplectic map description is eased by first
writing the SWM
as a full $N+1$ Hamiltonian system in the particle coordinates $(x_j,u_j)$
and the mean field
degrees of freedom \cite{tennyson,diego_CHAOS_00}
\begin{equation}
\frac{dx_{k}}{dt}=\frac{\partial {\cal H}}{\partial p_{k}}\,,\qquad \frac{
dp_{k}}{dt}=-\frac{\partial {\cal H}}{\partial x_{k}}\,,  \label{eq:1.17}
\end{equation}
\begin{equation}
\frac{d\theta }{dt}=\frac{\partial {\cal H}}{\partial J}\,,\qquad \frac{dJ}{
dt}=-\frac{\partial {\cal H}}{\partial \theta }\,, \label{eq:1.18}
\end{equation}
with Hamiltonian
\begin{equation}
\label{wp-hamiltonian}
{\cal H}=\sum_{j=1}^{N}\,\left[ \frac{1}{2\Gamma _{j}}\,p_{j}^{2}\,-2\Gamma
_{j}\,\sqrt{\frac{J}{N}}\,\cos (x_{j}-\theta )\right] -\Omega\, J\ .
\label{eq:1.19}
\end{equation}
  From (\ref{wp-hamiltonian}) it is clear that the SWM model consists
of $N$ {\it pendulum
Hamiltonians} mean-field-coupled through the amplitude $J$ and the
phase $\theta$.
Therefore, being the standard map the  symplectic discretization of the
pendulum Hamiltonian, the models studied here will consist of  an
ensemble of $N$ standard maps.
In the absence of coupling, i.e. ignoring self-consistency,
$J$ and $\theta$ would be constant
and the parameters of the standard map would be fixed numbers.
%and $\kappa$ and $\theta$ fixed numbers.
However, when self-consistency is incorporated, $J$ and $\theta$
become dynamical variables (also described by symplectic maps), and
this leads to a dependence of the parameters of the maps on
$\{x_1, x_2,\ldots x_N\}$ which gives rise to a global coupling of
the maps. The specific form of this coupling  will be discussed in
Section~\ref{sec:3}, where we present a systematic
discussion of the map models in terms of generating functions.

The remaining of this paper is organized as follows.
In Section \ref{sec:2} we briefly review the diffusion properties
in the case of the passive scalar, with particular emphasis on
the standard map. As mentioned before, in Section \ref{sec:3} we introduce
the two self-consistent systems studied in this work.
Sections~\ref{sec4} and \ref{sec5} are devoted
to the discussion of the numerical results. Section~\ref{sec6}
contains the  conclusions.

%%%%%%%%%%%%%%%%%%%%%%%%%%%%%%%%%%%%%%%%%%%%%%%%%%%%%%%%%%%%%%%%%%%%%%%
\section{A brief review of the diffusion properties of passive scalars}
\label{sec:2}

There exists a huge literature about the transport properties in the passive
scalar limit~\cite{moffat,ottino,bjpv}.
On the contrary, there are very few attempts in the study of the 
self-consistent
diffusion. The aim of this section is to recall the main results on
the diffusion problem for passive scalars in order to compare
them with the self-consistent diffusion that will be considered in the next
sections.
It is remarkable that the Lagrangian motion
can exhibit nontrivial behavior even for a
very simple velocity field ${\bf v}({\bf x},t)$~\cite{bjpv,aref}.
Complex behavior can be originated both from chaotic advection
(which is in general possible for stationary $3d$ flows or for
time-dependent $2d$ flows)
and/or from combined effects of the molecular
diffusivity and the advection velocity.
Under very general conditions (see below) the large scale field
$\langle \Theta \rangle$, which is obtained as an average of  a field
$\Theta$ evolving through Eq.~(\ref{eq:1.1})
on a volume whose dimensions are much larger than the typical length
scale of $\bv$, obeys at large times a diffusion equation
\BE
\frac{\partial \langle\Theta\rangle}{\partial t}=
\sum_{i,j} D_{ij}^E \frac{\partial^2 \langle \Theta \rangle}
{\partial x_i \partial x_j},
\label{eq:2.0}
\EE
with eddy diffusivity $D_{ij}^E$. In other words, the effect of the
velocity field  at large scales and time is the
renormalization of the transport coefficient $D_0$.
It is easy to understand the origin of (\ref{eq:2.0}) in the
Lagrangian framework.
Starting from (\ref{eq:1.6}), taking the average over many tracers, one has
\begin{eqnarray}
\langle \left( x_i(t)-x_i(0)\right)^2 \rangle
&=& 2 D_0 t+\int_0^t \int_0^t dt_1 dt_2
\langle v_i(\bx(t))v_i(\bx(t)) \rangle \nonumber \\
&=& 2 D_0 t + 2 \int_0^t dt_2 \int_0^{t_2} dt_1 C_{ii}(t_2-t_1),
\label{eq:2.1}
\end{eqnarray}
where we have assumed that $\langle v_i\left( \bx(t)\right)\rangle=0$
and we have introduced the correlation of Lagrangian velocities
$C_{ij}(t)\equiv \langle v_i(\bx(t)) v_j(\bx(0)) \rangle$.
At large times, if the correlation decays sufficiently fast,
the integral in (\ref{eq:2.1}) converges to an asymptotic value
\begin{equation}
\int_0^{\infty} dt C_{ii}(t)= \langle v_i^2 \rangle T_{L},
\label{eq:2.2}
\end{equation}
which defines the Lagrangian correlation time $T_{L}$.
  From (\ref{eq:2.1}) one recovers the Taylor result~\cite{taylor}
\begin{equation}
\langle \left( x_i(t)-x_i(0)\right)^2 \rangle
\approx 2 \left( D_0 + \langle v_i^2 \rangle T_{L} \right) t
\equiv 2 D_{ii}^{E} t,
\label{eq:2.3}
\end{equation}
which defines the eddy diffusivity in Eq.~(\ref{eq:2.0}).

Beyond the above typical scenario one can have anomalous dispersion, i.e.
\BE
\langle \left(x_i(t)-x_i(0)\right)^2 \rangle \sim  t^{2\nu},
\label{eq:2.4}
\EE
with $\nu \neq 1/2$. The case $\nu < 1/2$, called subdiffusion.
%is possible only in compressible flow.
Superdiffusion ($\nu > 1/2$) has been observed in incompressible
flows~\cite{diego_pof_98}, random shear flows and, as we will see, 
also symplectic
maps~\cite{zaslavsky,LL}. Anomalous diffusion can occur only if some 
of the hypothesis of
the above argument breaks down. Practically, this can be due to
two different mechanisms:
\begin{enumerate}
\item[(a)] Infinite variance of the velocity: $\langle v^2 \rangle = \infty$.
\item[(b)] Lack of decorrelation: $T_{L}=\infty$.
\end{enumerate}

The first condition, which leads to the class of L\'evy flights,
is not particularly realistic in physical systems, because it requires
infinite energy. We will not discuss here this behavior.
Case (b) is physically more relevant.
  From (\ref{eq:2.2}) one sees that anomalous superdiffusion is possible
only if $C_{ii}(t)$ goes to zero slower than $t^{-1}$. Unfortunately, the
behavior of $C_{ij}(t)$ is generated by the Lagrangian dynamics itself so
it is not trivial at all, in the absence of molecular diffusivity,
to determine whether the diffusion process will be standard or anomalous.
If molecular diffusivity is present  rather general results due to
  Avellaneda, Majda and Vergassola~\cite{maj} show that, if
the infrared contributions to $\bf v (\bf x,t)$ are not too strong, standard
diffusion occurs.

Let us now discuss the well known results for the diffusive behavior of the
standard map (a complete overview can be found in~\cite{LL}),
\begin{eqnarray}
x(t+1)&=&x(t)+y(t+1) \ \ mod \ 2\pi,
\label{eq:2.5} \\
y(t+1)&=&y(t)+K \sin(x(t)) \, .
\label{eq:2.6}
\end{eqnarray}
The Taylor argument, when applied to the $y(t)$ component (in this work
we always refer to the diffusion properties of $y(t)$)
of the standard map gives
\BE
D_y^E(K)=\frac{1}{2} K^2 \langle \sin^2 x \rangle
+\sum_{t=1}^\infty K^2 \langle \sin x(t) \sin x(0) \rangle \, .
\label{eq:2.7}
\EE
At large $K$, the map (\ref{eq:2.5}-\ref{eq:2.6}) exhibits widespread 
stochasticity
and to a good approximation consecutive angles
$x$ are decorrelated, thus one can neglect the second term in
(\ref{eq:2.7}) to obtain the quasi-linear (or random phase approximation,
RPA) result \cite{LL}:
\BE
D_y^E(K>>1)\approx D_{QL}=\frac{K^2}{4}.
\label{eq:2.8}
\EE
The above estimate is very crude: indeed it provides a good estimation
of the diffusion coefficient only at very high $K$.
Higher order corrections
to the RPA approximation can be obtained by means of the Fourier
technique \cite{LL}. At order $K^{-1/2}$ one obtains
\BE
D_y^E(K)={K^2 \over 4} \left[
   1-\sqrt{{8 \over \pi K}} \cos(K-\frac{5\pi}{4}) \right] \, .
\label{eq:2.9}
\EE
This approximation is rather good apart for small $K$ and around
{\it particular} values of $K$.
For $K \lesssim K_c \approx 0.972$ because of the presence of
separating KAM tori there is not diffusion at all,  $D_y^E(K)=0$.
On the other hand, at specific  values of $K$ (e.g. $K\approx 6.9115$)
corresponding to the
existence of ballistic solutions in the $y$ direction, instead of the
standard diffusion one observes an anomalous transport
with $\nu >1/2$~\cite{castiglione}.
%Let us remark that anomalous diffusion is a rare event, in the sense
%that it appears only at very precise values of $K$. Nevertheless,
%around these values, the diffusion coefficient $D_y^E$ is much larger
%than the approximate prediction (\ref{eq:2.9}).

%In Fig.~\ref{fig:difSM} we report the numerically evaluated
%$D_y^E$ normalized with the quasi-linear prediction $D_{QL}$
%in a large range of $K$. One can observe the slow convergence
%to $D_{QL}$  at large $K$. In the inset we
%plot the behavior of the diffusivity as a function of time in a
%neighborhood of the
%anomaly $K=6.9115$. The diffusivity is calculated as
%\begin{equation}
%D_y(t) = {\langle (y(t) -  y(0))^2 \rangle
%\over 2 t},
%\label{timedif}
%\end{equation}
%and the following anomalous behavior is observed: $D_y(t) ~ t^{}$.

%%%%%%%%%%%%%%%%%%%%%%%%%%%%%%%%%%%%%%%%%%%%%%%%%%%%%%%%%%%%%%%%%%%%%%%%
\section{Self-consistent map models}
\label{sec:3}

In this section we introduce the symplectic map models that we propose
for studying diffusion in self-consistent systems. As discussed in
Section~\ref{sec:1}, these maps consists of
ensembles of globally coupled standard maps.

The  definition of the
maps and the coupling  is guided by the well-known fact that if
$({\bf q}, {\bf p})$ denotes the canonical conjugate coordinates of a
  Hamiltonian system at time $n$, then  the transformation
$({\bf q}, {\bf p})\rightarrow ({\bf q}', {\bf p}')$  given by
\begin{equation}
\label{eq_34}
{\bf q}'=\frac{\partial S}{\partial {\bf p}'}\, , \qquad
{\bf p}=\frac{\partial S}{\partial {\bf q}}
\end{equation}
defines a symplectic map with generating function $S=S({\bf q},{\bf 
p}')$ \cite{LL}.

The generating functions of the models proposed here have the generic form:
\begin{equation}
S = S_p + S_f + S_i\, ,
\end{equation}
where $S_p$ defines the uncoupled evolution of the particles,
$S_f$ defines the uncoupled evolution of the mean field, and $S_i$ defines the
particles-mean field interaction.

For $S_p$ we assume the standard map generating function
\begin{equation}
S_p=\sum_{n=1}^N\left( x_n y_n' + \frac{1}{2 \Gamma_n}\, y_n'^2 + K_n
\cos x_n \right) \, ,
\end{equation}
where $\Gamma_n$, and $K_n$ are constants, the index $n$ labels
the particles, and we have used the notation
$x(t)=x$ and $x(t+1)=x'$.  In $S_p$ the particles are
uncoupled and each one follows independently a standard map dynamics.

In order to preserve the symplectic structure of the system, the field is
represented by two conjugate variables a phase $\theta$ and an
amplitude $J$ which,
in the absence of interaction with particles, evolve according to the
generating function
\begin{equation}
S_f=\theta J' + \int \omega(J')\, d J' \, ,
\label{eq:3.4}
\end{equation}
according to which
\begin{equation}
\left\{
\begin{array}{lll}
\theta' &=& \theta + \omega(J'), \\
J' &=& J.
\end{array}
\right.
\label{eq:3.3}
\end{equation}
That is, in the absence of coupling the amplitude and
frequency of
the mean field are constant.

The self-consistent coupling between the particles  and the mean field is
specified by two  functions $f$ and $g$ in the interaction generating function
\begin{equation}
%S(\{x_{n}\},\{y'_{n}\},\theta,J')=
S_i=
  g(J')\, \sum_{n=1}^{N} f(x_{n}-\theta)\, .
\label{eq:3.5}
\end{equation}
Based on this generating function, we will consider two models, one
introducing a
coupling only through the phase of the mean field, and another
introducing a
coupling through the phase and the amplitude of the mean field.

\subsection{Phase coupling}

In this case,  it is assumed that
\begin{equation}
g=\varepsilon\, ,  \qquad f=\sin (x_n-\theta)
\qquad \omega=\Omega\,J'\,  ,
\end{equation}
and $\Gamma_n=1$, $K_n=K$ for $n=1, 2 \ldots N$; with $\varepsilon$,
   $\Omega$,
$K$ constants.
The complete equations of motion thus become
\begin{equation}
\left\{
\begin{array}{lll}
x_{n}' &=& x_n + y'_n  \ \ \mbox{mod} \, 2 \pi, \\
y_{n}' &=& y_n + K \sin x_n + \varepsilon \cos(x_n-\theta), \\
\theta' &=& \theta + \Omega\,  J', \\
J' &=& J - \varepsilon \sum_{n} \cos (x_n-\theta). \\
\end{array}
\right.
\label{eq:3.6}
\end{equation}
The parameter $\varepsilon$ measures the strength of the
coupling. For $\varepsilon=0$ we recover the situation discussed
in the previous Section for $N$ independent particles.
One may expect that, in general, the inclusion of a coupling
between the different particles will change the diffusive
behavior of the system.
If there were no feedback from the particle variables $x_n$ to
the field $J$ (i.e. no self-consistency) one could argue
that the effect of the field variable $\theta$ in (\ref{eq:3.6})
would be the same as that of a noise and thus will destroy the correlations
in the $y_n$ variables.
In this case one expects that the deviations of the diffusion
coefficient with respect to the quasilinear prediction discussed in the
previous Section would be strongly suppressed.
Of course, this kind of argument can not be completely
justified in presence
of the full coupling in (\ref{eq:3.6}).
 
%%%%%%%%%%%%%%%%%%%%%%%%%%%%

\subsection{Amplitude and phase coupling}

In this case,  it is assumed that
\begin{equation}
g=2 \sqrt{J'} \, ,  \qquad f=\Gamma_n \cos (x_n-\theta) \, ,
\qquad \omega=- \Omega \, ,
\end{equation}
where $\Omega$ is constant and $\Gamma_n$ are $
N$ independent parameters.  The dynamics in this case is determined by
\begin{equation}
\left\{
\begin{array}{lll}
x'_n &=& x_n +  y'_n/\Gamma_n, \, \\ \nonumber
y'_n &=& y_n +K\, \sin x_n - 2\,   \Gamma_n \sqrt{J'}\, \sin (
x_n-\theta), \, \\ \nonumber
%\label{map-11}
\theta' &=& \theta  -\Omega -\frac{1}{\sqrt{J'}} \,
\sum_{n=1}^N\, \Gamma_n \cos(x_n - \theta), \, \\ \nonumber
J' &=& J + 2 \,\sqrt{J'} \,
\sum_{n=1}^N\, \Gamma_n \sin(x_n - \theta) \ ,
\end{array}
\right.
\label{eq:full}
\end{equation}
where, to simplify matters, we have assumed that the external field
is such that
$K_n/\Gamma_n=K$ being $K$  a constant.
This  globally coupled map was originally proposed in 
Ref.~\cite{diego_CHAOS_00}
as a  symplectic discretization of
the single wave model Hamiltonian system in Eq.~(\ref{wp-hamiltonian}).
Compared with (\ref{eq:3.6}), in the map  (\ref{eq:full})  $J$ and 
$\theta$ are both
coupled to the particles and this leads to a self-consistent
modification of the phase {\em
and} the amplitude of the mean field.

The map for $J$   is implicit. However,
   rescaling  the variables $y_n \rightarrow \Gamma_n \,y_n$,
$K\rightarrow \Gamma_n \,K$, and defining
\begin{equation}
\kappa=2 \sqrt{J}\, , \qquad
   \gamma_n = 2 \, \Gamma_n \, , \qquad
\eta=  \sum_{n=1}^N \gamma_n \sin \left(  x_n - \theta \right) \ ,
\end{equation}
the map can be written in a fully explicit form as~\cite{diego_CHAOS_00}
\begin{equation}
\left\{
\begin{array}{lll}
x_n' &=& x_n + y_n' \ ,\\ \nonumber
%\label{map-17}
y_n' &=& y_n + K\, \sin x_n -
\kappa' \, \sin \left(x_n-\theta\right) \ ,  \\ \nonumber
%\label{map-18}
\kappa' &=& \sqrt{ \kappa^2 +  \eta^2}
+ \eta \ ,  \\ \nonumber
%\label{map-19}
\theta' &=& \theta - \Omega +
\,\frac{1}{\kappa'}\, \frac{\partial\, \eta}{\partial \theta} \ .
\end{array}
\right.
\label{D_map}
\end{equation}
In this model, the mean field amplitude, $\kappa$,  plays the role of the
standard map parameter (besides the $K$ constant)
which is self-consistently coupled to the particles through
the order parameter
$\eta$.
In the mean field particle dynamics one can define the total
{\it momentum} of the  system as
%\BE
%{\cal P}=\sum_{n=1}^N\, y_n + J \, ,
%\EE
\BE
{\cal P}=\frac{ \kappa^2}{2}+\sum_{n=1}^N
   \gamma_n \, y_n \, ,
\label{momentum}
\EE
where the first term on the right hand side represents the
momentum of the mean field,
and the second term the total momentum of the  particles . This quantity
is a constant of motion of  (\ref{eq:full}),
\BE
{\cal P}'={\cal P}\, .
\EE
This conservation law plays an important role in the diffusive 
properties of the system.

%%%%%%%%%%%%%%%%%%%%%%%%%%%%%%%%%%%%%%%%%%%%%%%%%%%%%%%%%%%%%%%%%%%%%%
\section{Diffusion in phase coupled maps}
\label{sec4}

Let us study the diffusion properties in the phase coupled map
(\ref{eq:3.6}). In
particular, we will see that the effect of the mean
field coupling is the randomization of
  the phase. That is, when coupling occurs only through the phase, 
self-consistency
increases the stochasticity of the map, and the diffusive properties
of the coupled maps are practically indistinguishable from the dynamics of
a phase-randomized uncoupled map.

In the limit $\varepsilon=0$ the diffusion coefficient (for the $y$ component),
as a function of the parameter $K$,
displays the complex behavior discussed in Section~\ref{sec:2}.
The first natural question is whether this complexity survives
in the presence of  coupling with a self-consistent
mean field (i.e. for $\varepsilon>0$).

\begin{figure}
\centerline{\epsfig{figure=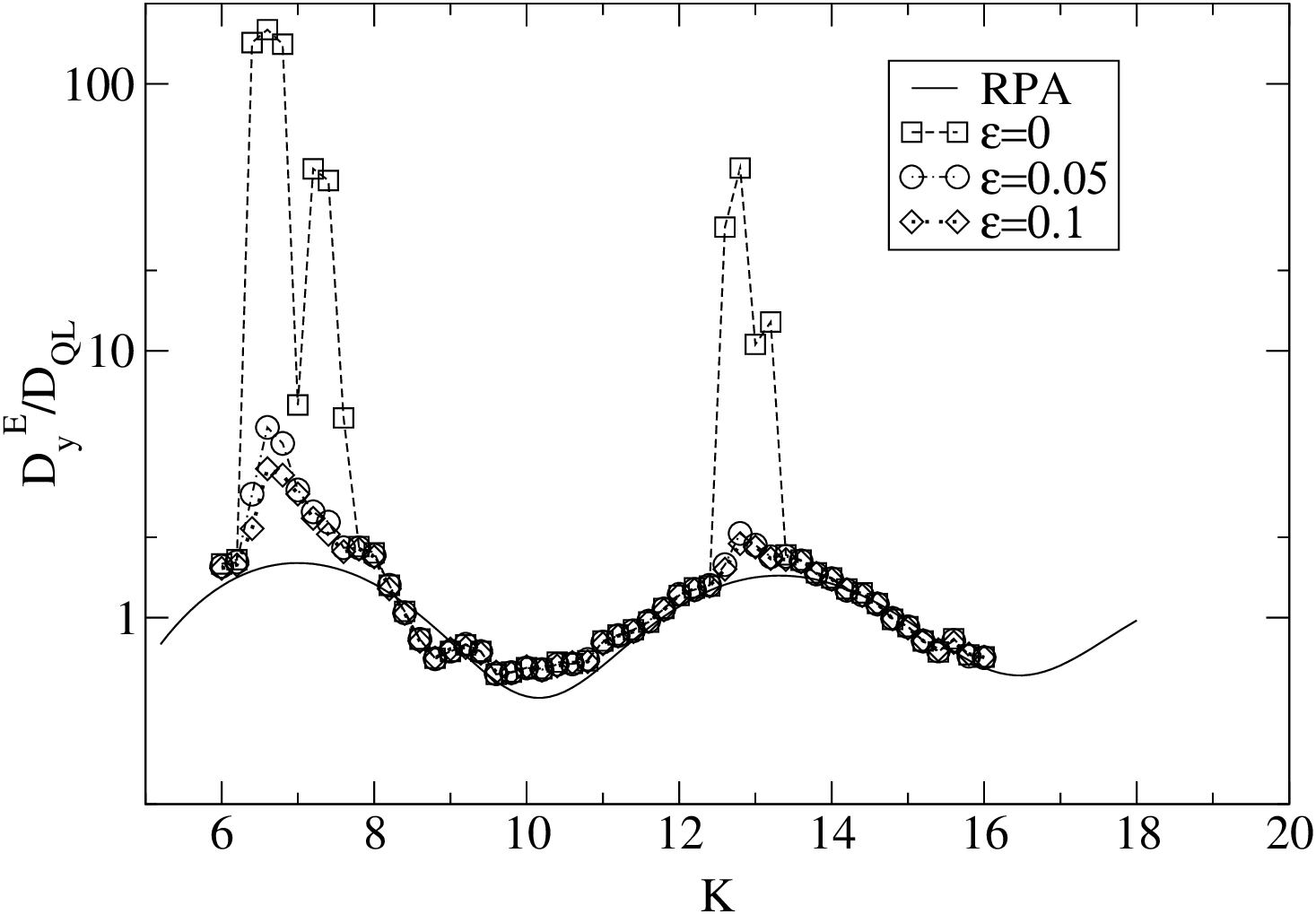,width=\columnwidth,angle=0}}
\caption{Normalized diffusion coefficient vs. $K$ for different values of the
$\varepsilon$ parameter; squares for $\varepsilon=0$, circles
$\varepsilon=0.05$ and diamonds for  $\varepsilon=0.1$.
Solid line represents the theoretical prediction as given by the
higher order RPA approximation. The number of particles is $N=60000$,
$\Omega=0.1$ and the number of steps is $10000$.}
\label{fig:prin}
\end{figure}

In Fig.~\ref{fig:prin} we show $D_y^E(K)$ (normalized with the RPA
prediction $K^2/4$) in the uncoupled case ($\varepsilon=0$) and
for some other values of the  coupling ($\varepsilon=0.05$ and
$\varepsilon=0.1$).
Numerically, $D_y^E(K)$ is calculated by taking the large time limit
of the expression:
\begin{equation}
D^E_y(t) = {\langle (y(t) -  y(0))^2 \rangle
\over 2 t}.
\label{timedif}
\end{equation}
   For comparisons, we also plot the value of
$D_y^E(K)/D_{QL}$ with the higher order RPA corrections as given
by (\ref{eq:2.9}).
As one can see, for a generic large value of $K$ the presence of
a small coupling do not change the diffusion coefficient, apart for
the values at which one can observe anomalous
diffusion in the uncoupled limit ($\varepsilon=0$).
Therefore, the first effect of the mean
field is to remove the ballistic contributions to the dispersion.

\vspace{3cm}
%%%%%%%%%%%%%%%%%%%%%%%%%%%%%%%%%%%%%%%%%%%%%%%%%%%%%%%%%%%%%%%%%%%%%%
%%%%%%%%%%%%
%%%%%%%%%%%%%%%%%%%%%%%%%%%%%%%%%%%%%%%%%%%%%%%%%%%%%%%%%%%%%%%%%%%%%%
%%%%%%%%%%%

Thus, let us try to focus on the case where the standard map ($\varepsilon=0$)
shows an anomalous behavior, for e.g. $K=6.9115$, and see
where the differences with the
self-consistent map ($\varepsilon \ne 0$) appear.
The diffusion coefficient is an asymptotic quantity. For
finite time, the evolution of Eq.~(\ref{eq:3.6}) with small
$\varepsilon$ maintains a memory of the behavior
at $\varepsilon=0$ up to a time $T(\varepsilon)$ (saturation time).
We define $T(\varepsilon)$ as the time at which the
finite-time diffusion coefficient, as given by Eq.~(\ref{timedif}),
is reasonably close to its  asymptotic value.
 
\begin{figure}
\epsfig{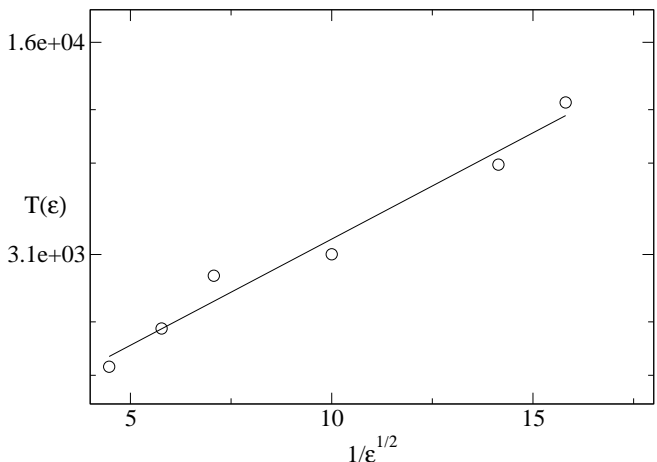}
\caption{Saturation time, $T(\varepsilon)$, vs. $1/\sqrt{\varepsilon}$
  for $K=6.9115$ and $N=60000$.}
\label{fig:Tsat}
\end{figure}

\begin{figure}
\centerline{\epsfig{figure=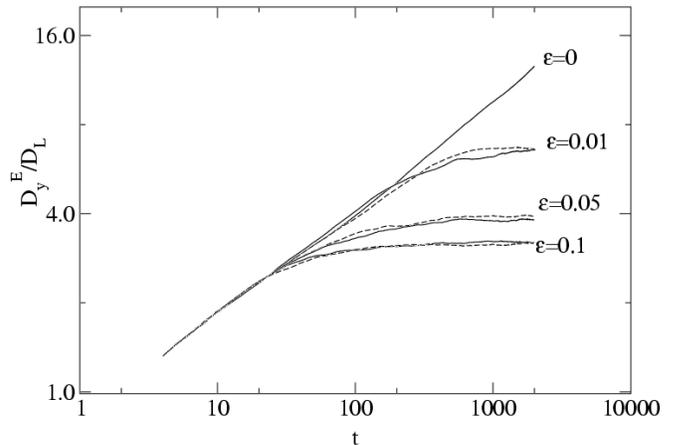,width=\columnwidth,angle=0}}
\caption{$D^E_y/D_L$ against time  for two different number of particles,
   $K=6.9115$, and different values of $\varepsilon$. Solid line is for
$N=60000$ and the dashed-line for $N=40000$. }
\label{fig:Ndep}
\end{figure}

In Fig.~\ref{fig:Tsat} we show $T(\varepsilon)$,
for a system with $K=6.9115$, as a function of $1/\sqrt{\varepsilon}$.
The approximately linear behavior in the plot indicates
the dependence $T(\varepsilon) \sim \exp(c/\sqrt{\varepsilon})$, with
$c$ an arbitrary constant.
Let us note that the anomalous diffusion is mainly
due to the presence of ballistic (non-chaotic) trajectories. Therefore the
failing of anomalous transport can be seen as the recover of a generic
statistical behavior.
In this sense this is consistent with a scenario {\it \'a la
Nekhoroshev}~\cite{LL}.
We also remark that, within the range of values investigated,
there is not evidence of dependence on the size $N$ as can be
seen in Fig.~\ref{fig:Ndep} (again for $K=6.9115$).
  It is worth mentioning here that
in  Fig.~\ref{fig:Ndep} for $\varepsilon=0$ one observes the
typical anomalous behavior of the diffusivity, that is,
$D^E_y \propto t^{0.3}$~\cite{castiglione}.

\vspace{3cm}
%%%%%%%%%%%%%%%%%%%%%%%%%%%%%%%%%%%%%%%%%%%%%%%%%%%%%%%%%%%%%%%%%%%%%%%%%%%%%%
%%%%%%%%%%%%%%%%%%%%%%%%%%%%%%%%%%%%%%%%%%%%%%%%%%%%%%%%%%%%%%%%%%%%%%%%%%%%%

%%%%%%%%%%%%%%%%%%%%%%%%%%%%%%%%%%%%%%%%%%%%%%%%%%%%%%%%%%%%%%%%%%%%%%%%%%%%%%
%%%%%%%%%%%%%%%%%%%%%%%%%%%%%%%%%%%%%%%%%%%%%%%%%%%%%%%%%%%%%%%%%%%%%%%%%%%%%
 
\begin{figure}
\epsfig{figure=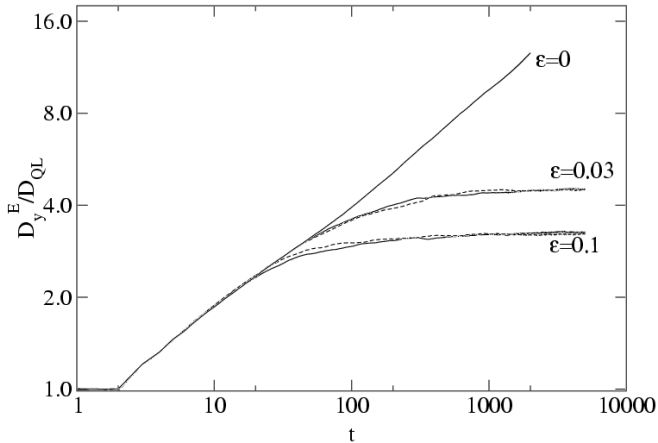,width=\columnwidth,angle=0}
\caption{ $D_y^E(K=6.9115)/D_{QL}$ vs. time for different values of 
$\varepsilon$
and $N=60000$. The solid-line is for
an external self-consistent field and the dashed-line is for an
external random field.
The random external field is generated by picking,
at every time step, a random number regularly distributed
in the interval $[ 0, 2\pi]$.
}
\label{fig:random}
\end{figure}

The discussed results indicate that the main effect of the
self-consistent field is to reduce the deviations from the
statistical prediction.
   This is to be expected because in this case, as mentioned
before, the phase coupling
leads to a randomization of the phase which is precisely what is
assumed in the statistical
arguments based on the random phase approximation.
We check this statement by replacing the self-consistent field
with an external noise. We study a system of
$N$ particles whose evolution is now given by a time-dependent
generating function
\begin{equation}
S(\{x'_{n}\},\{y_{n}\},t)=
\sum_{n=1}^{N} S_0(x'_{n},y_{n}) +
\varepsilon \sum_{n=1}^{N} f(x'_{n}-\eta(t))
\label{eq:4.2}
\end{equation}
where $\eta$ is a random process with the same statistical properties
of the self-consistent $\theta$, i.e., $\eta$ is a random
number uniformly distributed in the interval $[0, 2\pi ]$.
The result  for $D_y^E (K=6.9115)/D_L$ against time is plotted in
Fig.~\ref{fig:random}
for both the self-consistent and the random field.
One observes that the random approximation is rather accurate.
Therefore, taking into account the results for $K$ large and the
specific values of $K$ where the standard map is anomalous, one can say that
the dynamics for $K$ large in the self-consistent map model
(\ref{eq:3.6}) is equivalent to an {\it effective}
standard map.

%\vspace{2cm}
%%%%%%%%%%%%%%%%%%%%%%%%%%%%%%%%%%%%%%%%%%%%%%%%%%%%%%%%%%%%%%%%%%%%%%
%%%%%%%%%%
%%%%%%%%%%%%%%%%%%%%%%%%%%%%%%%%%%%%%%%%%%%%%%%%%%%%%%%%%%%%%%%%%%%%%%
%%%%%%%%%%%

\begin{figure}
\epsfig{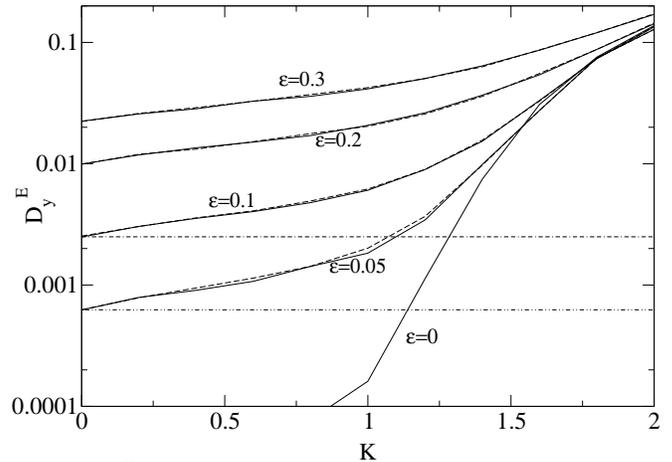}
\caption{ $D_y^E$ vs $K$  for different values of $\varepsilon$. Here the
number of particles is $N=60000$ and the final time is $10000$.
The solid-line is for
an external self-consistent field and the dashed-line is for an
external random field.
The straight lines correspond to the lines $D_y^E=\varepsilon^2/4$ for
$\varepsilon=0.1$ (dashed-dotted) and $\varepsilon=0.05$ (dashed-doubledotted)
}
\label{fig:randomKpi}
\end{figure}

Most interesting, a priori,  is the case of small values of $K$.
As we have recalled in the previous Section, for $K <1$
and $\varepsilon=0$ there is no diffusion due to the  presence of KAM
tori. However, as Fig.~\ref{fig:randomKpi} shows, the phase coupled
($\varepsilon \neq 0$) self-consistent map displays finite
diffusion for arbitrary small values of $K$. This is, once again, a
manifestation of self-consistent driven phase randomization.

\begin{figure}
\epsfig{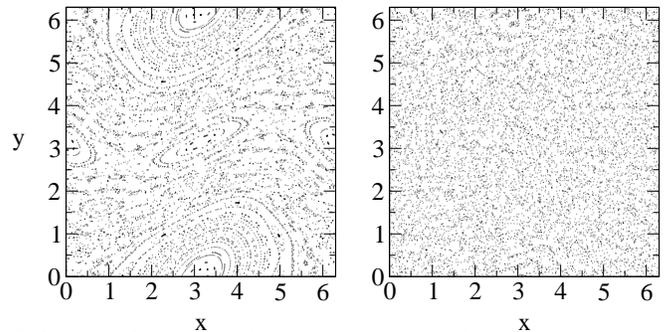}
\caption{Left, standard map phase space for $K=1$, and, right,
the trajectory of one particle for the system
(\ref{eq:3.6}) with $K=1$ and $\varepsilon=0.1$.}
\label{fig:phasespace}
\end{figure}

The same
scenario as for large values of $K$ has been identified, that is, the
external self-consistent field is equivalent, when diffusion properties are
under study, to an external random field. This can be seen again in
Figure~\ref{fig:randomKpi} where we also plot the diffusion coefficient
obtained from the effective random map.
At small $K$, one observes a weak dependence of $D_y^E$ on $K$,
while $D_y^E \simeq \varepsilon^2/4$, leading to a finite diffusivity
also for $K=0$.
As mentioned before, it is the breaking of the regular orbits
of the standard map (for $K$ small) induced by the mean field, that allows the
diffusion of particles. This is clearly seen  in
Figure~(\ref{fig:phasespace}) where it is  shown some trajectories in 
the phase
space for $K=1$ for both the standard map and the
coupled map model (\ref{eq:3.6}) with $\varepsilon=0.1$.

In summary, for model (\ref{eq:3.6}) the coupling to an external
self-consistent mean field is equivalent to the effect induced
by an external random field.
This result is valid also for $K<1$ where
the standard map shows barriers to transport.

\vspace{2cm}
%%%%%%%%%%%%%%%%%%%%%%%%%%%%%%%%%%%%%%%%%%%%%%%%%%%%%%%%%%%%%%%%%%%%%%
%%%%%%%%%%%%%%%%

%%%%%%%%%%%%%%%%%%%%%%%%%%%%%%%%%%%%%%%%%%%%%%%%%%%%%%%%%%%%%%%%%%%%%%
%%%%%%%%%%%%%%

\vspace{3cm}
%%%%%%%%%%%%%%%%%%%%%%%%%%%%%%%%%%%%%%%%%%%%%%%%%%%%%%%%%%%%%%%%%%%%%%
%%%%%%%%%%%%%%%%

%%%%%%%%%%%%%%%%%%%%%%%%%%%%%%%%%%%%%%%%%%%%%%%%%%%%%%%%%%%%%%%%%%%%%%
%%%%%%%%%%%%%%

\section{Diffusion in fully self-consistent maps}
\label{sec5}

In this section we study the diffusive properties of fully
self-consistent maps. By this we mean maps coupled through the
phase {\em and} the amplitude of the mean field.
Our study will be based on the map (\ref{D_map}) with $\Omega=0$, and in
the absence of external
field ($K=0$) since we are mostly interested in self-consistent effects.

We consider an ensemble of $N$ particles with coordinates
$(x_j, y_j)$, $j=1, 2\ldots N$, in the rectangle
$x \in (-\pi, \pi)$ and $y \in (-\pi/2, \pi/2)$.
We will focus in the study of a Gaussian distributed active field
\BE
\gamma_j=\gamma_0\exp\left[\frac{-(x_j^2+y_j^2)}{2\sigma^2}\right] \, .
\label{gaussian}
\EE
For these nonuniform $\gamma_j$ distributions it is useful to distinguish
between the particle
variance $\sigma^2_{p y}$, and the concentration
variance $\sigma^2_{\gamma y}$ defined as
\BE
\sigma^2_{p y}=\langle \left[y - \langle y \rangle_p \right]^2
\rangle_p \, , \qquad
\sigma^2_{\gamma y}=\langle \left[y - \langle y \rangle_\gamma
\right]^2 \rangle_\gamma \, ,
\EE
where
\BE
\langle q \rangle_p=\frac{1}{N}\, \sum_{j=1}^N\, q_j\, , \qquad
\langle q \rangle_\gamma=\frac{1}{N}\,  \sum_{j=1}^N\, \gamma_j q_j\, .
\EE

\subsection{Subcritical diffusion}

\begin{figure}
\epsfig{figure=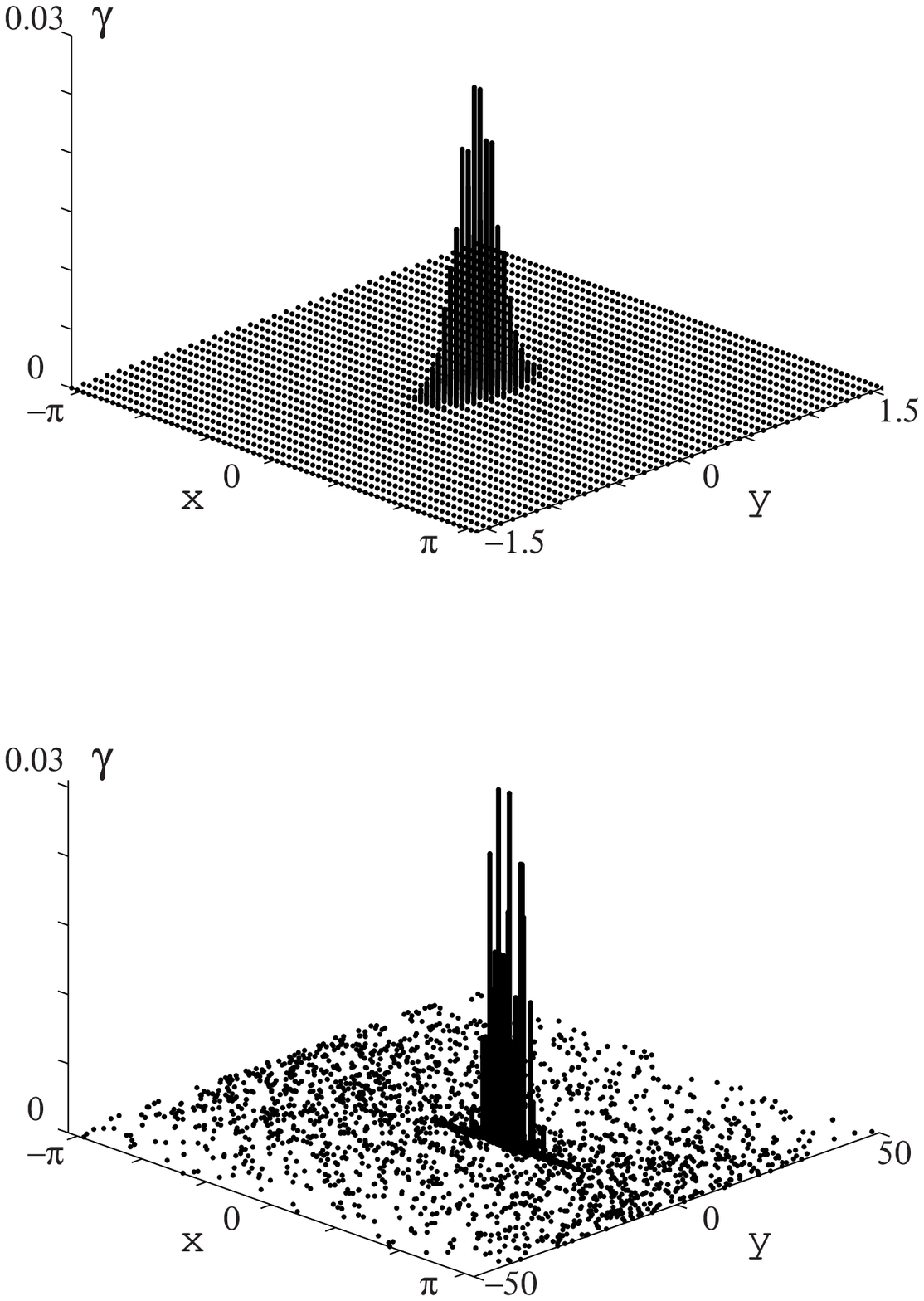,width=\columnwidth,angle=0}
\caption{Phase space particle distribution in the fully self-consistent map
(\ref{D_map}) for a Gaussian distributed active scalar according to
(\ref{gaussian}) with $2\sigma^2=0.2$ and $\gamma_0=0.0269$ and initial
conditions $\kappa(1)=0.8$ and $\theta(1)=0$. The two panels show the particle
distribution at the initial and final time, after 137000 steps.
The height of a vertical lines corresponds to the active
scalar concentration $\gamma_j$ of the $j$-th particle located at
$(x,y)=(x_j,y_j)$.}
\label{fig_cluster_below}
\end{figure}

\begin{figure}
\epsfig{figure=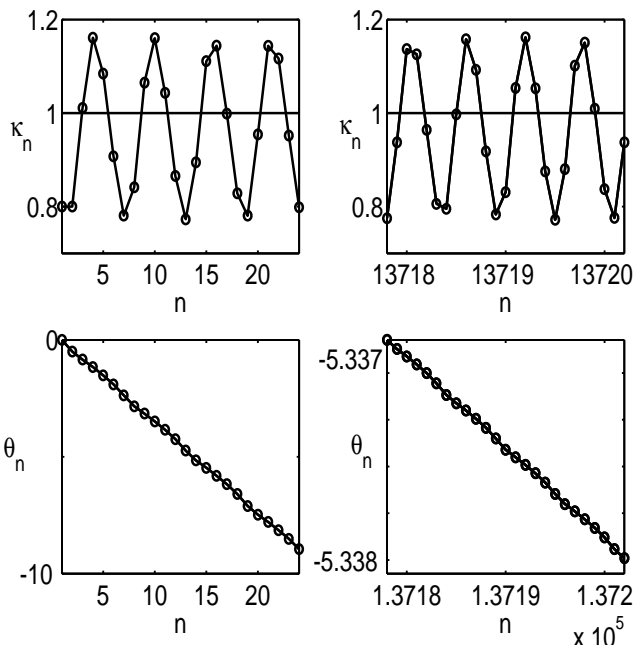,width=\columnwidth,angle=0}
\caption{ Time evolution of $\kappa$ and $\theta$  in the fully
self-consistent map
(\ref{D_map}) for a Gaussian distributed active scalar according to
Eq.~(\ref{gaussian}) and initial
conditions
$\kappa(1)=0.8$ and $\theta(1)=0$. The plots show the evolution in
time windows at
the beginning $n\in (0, 60)$ and at the end $n\in (137000, 137060)$
of the run. The same periodic
behavior is observed at intermediate times}
\label{fig_kappa_th_below}
\end{figure}

\begin{figure}
\epsfig{figure=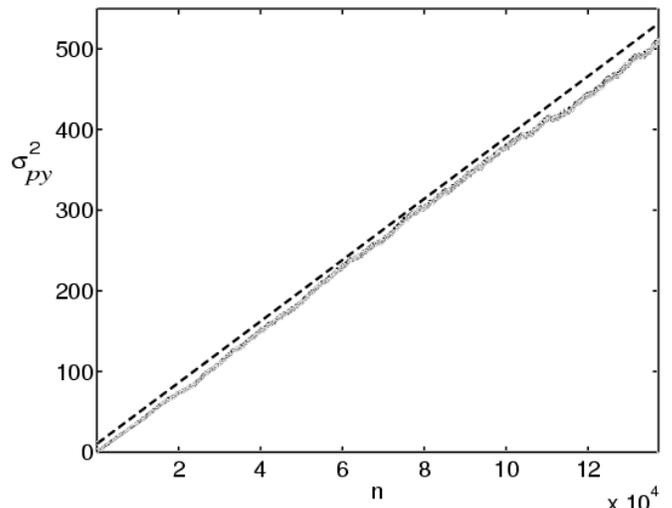,width=\columnwidth,angle=0}
\caption{
Subcritical diffusion
in the fully self-consistent map
(\ref{D_map}) for a Gaussian distributed active scalar according to
Eq.~(\ref{gaussian}) and initial
conditions
$\kappa(1)=0.8$ and $\theta(1)=0$. Even though in this case, as shown in
Fig.~\ref{fig_kappa_th_below}, the mean value of $\langle \kappa
\rangle$ is below the critical
value for the destruction of all KAM barriers, the variance shows
clear evidence of diffusive
transport with
   $2D_y^E=0.0038$.}
\label{fig_vari_below}
\end{figure}

In the first simulation we iterated the map with initial
conditions  $\kappa (1)=0.8$, and $\theta (1)=0$.
The upper panel of Fig.~\ref{fig_cluster_below} shows the initial $\gamma$
distribution.
For $t>0$, the scalar mixes and in the process modifies $\kappa$ and $\theta$.
In particular, in this case, as shown in
Fig.~\ref{fig_kappa_th_below}, $\kappa$ oscillates  in time around a
mean value $<\kappa>=0.966$ ($<.>$ is the temporal average)
slightly below the critical value $\kappa_c=0.9716$  for the
destruction of KAM barriers and the onset of diffusion in the standard map.
The phase (see also Fig.~\ref{fig_kappa_th_below}) decreases monotonically.
%%%%%%%%%%%%%%%%%%%%%%%%%%%%%%%%%%%%%%%%%%%%%%%%%%%%%%%%%%%%%%%%%%%%%%%%
%\vspace{2cm}
%%%%%%%%%%%%%%%%%%%%%%%%%%%%%%%%%%%%%%%%%%%%%%%%%%%%%%%%%%%%%%%%%%%%%%%%
That is, the self-consistent coupling drives the system periodically
between a diffusive regime with no KAM barriers ($\kappa>\kappa_c$)
and a non-diffusive regime
with KAM barriers ($\kappa<\kappa_c$).
As shown in Fig.~\ref{fig_vari_below}, this yields to
diffusive particle transport in $y$, $\sigma^2_{p y}= 2 D t$, even
though on the average  $\kappa$ is
below the threshold for diffusion, that is there is subcritical
diffusion.  Because the peak of the $\gamma$
distribution remains coherent, in this case there is no diffusion in
the concentration, i.e.  $\sigma^2_{\gamma y}=0$.
As discussed in the previous Section, also in the phase
coupled map (\ref{eq:3.6}) there is a subcritical diffusion regime.
However, there is an important difference between these two cases
because in the phase coupled map the time evolution of the phase
$\theta$ exhibits random behavior while in the fully self-consistent
map, as shown in Fig.~\ref{fig_kappa_th_below},
$\theta$ has a regular behavior.

The oscillations of $\kappa$ are caused by the feedback effect of
the active scalar trapped in the
period-one island of the map (see Fig.~\ref{fig_cluster_below}).
We describe the coherent part of the distribution, i.e. the part of
the distribution trapped in the
period-one island,  as  a {\it macroparticle}. The  macroparticle
representation is a sort of renormalization
process in which a group of particles with different values of
$\gamma_k$ are replaced by one with
an effective $\gamma$.  The macroparticle concept  provides a link between
systems with large (or infinite)  degrees of freedom  and low
dimensional systems~\cite{tennyson,diego_CHAOS_02,springer_02}.
In the case considered here,  at a given time $t$, the macroparticle
rotates around the o-point of an effective standard map with
coupling constant $K=\kappa(t)$ and phase $\theta(t)$.

The  oscillation period for a standard map of coupling parameter $K$ can
be estimated as \cite{LL}
\BE
%T=\frac{2 \pi }{\cos^{-1}\left(1-\frac{K}{2}\right)}
T=\frac{2 \pi }{ \arccos \left(1-\frac{K}{2}\right)} \, .
\label{period}
\EE
%%%%%%%%%%%%%%%%%%%%%%%%%%%%%%%%%%%%%%%%%%%%%%%%%%%%%%%%%%%%%%%%%%%%%%
%%%%%%%%%%%%%%
%\vspace{2cm}
%%%%%%%%%%%%%%%%%%%%%%%%%%%%%%%%%%%%%%%%%%%%%%%%%%%%%%%%%%%%%%%%%%%%%%
%%%%%%%%%%%%%%
%\vspace{2cm}
%%%%%%%%%%%%%%%%%%%%%%%%%%%%%%%%%%%%%%%%%%%%%%%%%%%%%%%%%%%%%%%%%%%%%%%%

To use this result to calculate the rotation period of the self-consistent
oscillation of $\kappa$, note that
according to  the conservation of momentum in (\ref{momentum}),
\BE
\kappa^2(t)=\kappa^2(1)+\sum_{n=1}^N\, \gamma_n\,
\left[y_n(1)-y_n(t)\right] \, .
\EE
In the macroparticle description, this relation can be written as
\BE
\label{macrop}
\kappa^2(t)= \kappa^2(1)+\Gamma_{mc}\, Y_{mc} (t)\, ,
\EE
where $\Gamma_{mc}=\sum_n \gamma_n$ is the effective $\gamma_{eff}$ of the
macroparticle and
$Y_{mc}$ the $y$-coordinate of the macroparticle.
According to (\ref{macrop}), the oscillation period of $\kappa$ equals the
rotation period of the macroparticle which can be estimated using
(\ref{period}) with $K=\langle
\kappa
\rangle$. For $\langle \kappa \rangle\approx 1$ this approximation
gives a period
$T \approx 6$ which is in good agreement with the numerical results
(see Fig.~{\ref{fig_kappa_th_below}).

%%%%%%%%%%%%%%%%%%%%%%%%%%%%%%%%%%%%%%%%%%%%%%%%%%%%%%%%%%%%%%%%%%%%%%%%
\subsection{Self-consistent suppression of diffusion}

In the previous example, the constant rotation of the active scalar
trapped in the period-one
island  gave rise to stationary oscillations of $\kappa$ and steady
particle  diffusion in $y$.
However, this is not always the case, and it is possible that
diffusion is suppressed rather than
maintained by  self-consistent effects.
As an example,  consider the same initial conditions as before but
with a smaller initial value
of the coupling parameter $\kappa$, namely $\kappa(1)=0.6$.  In this case,
as Fig.~\ref{fig_diff_suppress} shows, there is an initial regime in which
$\kappa$ oscillates beyond $\kappa_c$ and diffusive transport is present  with
$D=0.0014$.
However, after a  fraction of particles have migrated to regions of large $y$,
$\kappa$ drops systematically below $\kappa_c$ and diffusion is
suppressed.  At this point the
system enters in  a  transient subdiffusive regime leading to the
eventual elimination of the
diffusion.
As shown in Fig.~\ref{fig_diff_suppress}, and in more detail
in Fig.~\ref{fig_kappa_damp},  the suppression of the
diffusion is accompanied by a
damping of the coupling parameter $\kappa$. Note that consistent
with the estimation in
(\ref{period}), the period of oscillation remains constant $T\approx
6$. According to
momentum conservation in Eq.~(\ref{macrop}), this
damping can be viewed as a momentum transfer from the mean-field to
the particles. This is reminiscent of the Landau damping mechanism in 
plasmas in which an
energy transfer from the field to the particles leads to a 
colissionless damping of the
field.

%%%%%%%%%%%%%%%%%%%%%%%%%%%%%%%%%%%%%%%%%%%%%%%%%%%%%%%%%%%%%%%%%%%%%%
%%%%%%%%%%%%%%
%\vspace{2cm}
\begin{figure}
\epsfig{figure=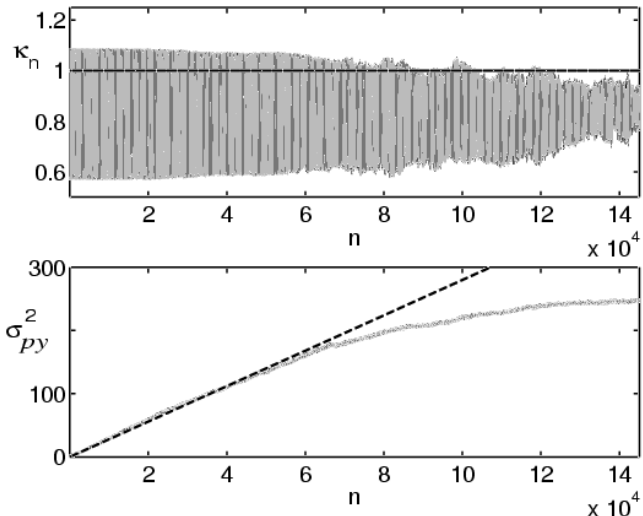,width=\columnwidth,angle=0}
\caption{Self-consistent suppression of diffusion in the
map (\ref{D_map}) for a Gaussian distributed active scalar according
to  Eq.~(\ref{gaussian}) and
initial conditions
$\kappa(1)=0.6$ and $\theta(1)=0$. Upper panel  shows the time evolution
of $\kappa$ and the lower panel the
time evolution of the square of the particle variance.  For $n \in
(1, 6\times 10^4)$, $\kappa$ reaches
values above the threshold for KAM barriers destruction ($\kappa_c
\approx 1$). For later times, the
maximum of $\kappa$ drops systematically below $\kappa_c$ and
diffusion is suppressed.
}
\label{fig_diff_suppress}
\end{figure}
%%%%%%%%%%%%%%%%%%%%%%%%%%%%%%%%%%%%%%%%%%%%%%%%%%%%%%%%%%%%%%%%%%%%%%
%%%%%%%%%%%%%%
%\vspace{2cm}
\begin{figure}
\epsfig{figure=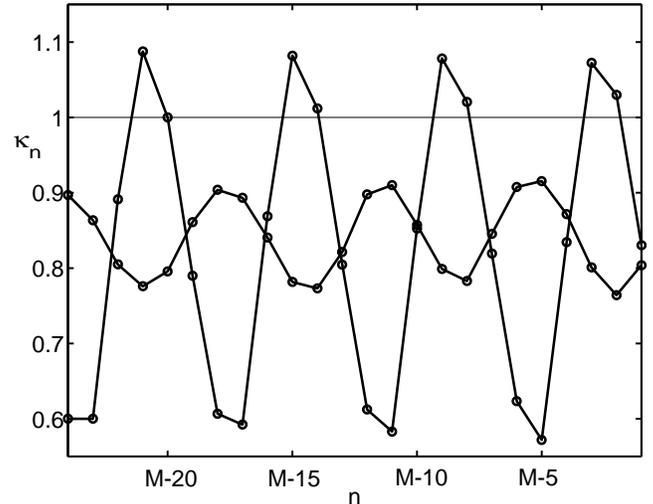,width=\columnwidth,angle=0}
\caption{Damping of $\kappa$ during the self-consistent suppression
of diffusion shown in
Fig.~\ref{fig_diff_suppress}. The curve with larger amplitude shows
the beginning of the time series
(top panel of Fig.~\ref{fig_diff_suppress} with $M=25$) and the curve
with the smaller amplitude
shows the tail of the time series (top panel of
Fig.~\ref{fig_diff_suppress} with
$M=14\times 10^4$).}
\label{fig_kappa_damp}
\end{figure}
%%%%%%%%%%%%%%%%%%%%%%%%%%%%%%%%%%%%%%%%%%%%%%%%%%%%%%%%%%%%%%%%%%%%%%%%

%%%%%%%%%%%%%%%%%%%%%%%%%%%%%%%%%%%%%%%%%%%%%%%%%%%%%%%%%%%%%%%%%%%%%%%%
\subsection{Macroparticle instability and diffusion of concentration}

In the previous examples, the value of $\kappa$ was relatively small,
and  the macroparticle (i.e. the conglomerate of particles with the
largest values of $\gamma$) remained coherent.
This yields to either  steady diffusion (see
Fig.~\ref{fig_vari_below}) or transient diffusion
(see Fig.~\ref{fig_diff_suppress}) of the
particle distribution, but no diffusion of the concentration, i.e.
$\sigma_{\gamma y}\approx 0$.
However, self-consistent effects can destabilize the macroparticle
and the concentration field diffuses.
To illustrate this, we consider the same Gaussian distribution
(Eq.~(\ref{gaussian}) as before but with a larger initial value of $\kappa$,
namely $\kappa (1)=3.3$). In this case,
as Fig.~\ref{fig_macropart_insta} shows,
up to $n\approx 5 \times 10^3$, $\kappa$ remains approximately constant,
the phase $\theta_n$ (panel (b)) decreases monotonically, the
concentration variance $\sigma_{\gamma y}$  does not grow and the
particle variance $\sigma_{p y}$ exhibits standard diffusion.
Around $n\approx 5 \times 10^3$ there is a transition and
$\kappa$ grows rapidly giving rise to a diffusion of the concentration
and a jump in the particle diffusion.
Figure~\ref{fig_macropart_insta_phase} shows the active scalar distribution
at two different times.
%%%%%%%%%%%%%%%%%%%%%%%%%%%%%%%%%%%%%%%%%%%%%%%%%%%%%%%%%%%%%%%%%%%%%%
%%%%%%%%%%%%%%
%\vspace{2cm}
\begin{figure}
\epsfig{figure=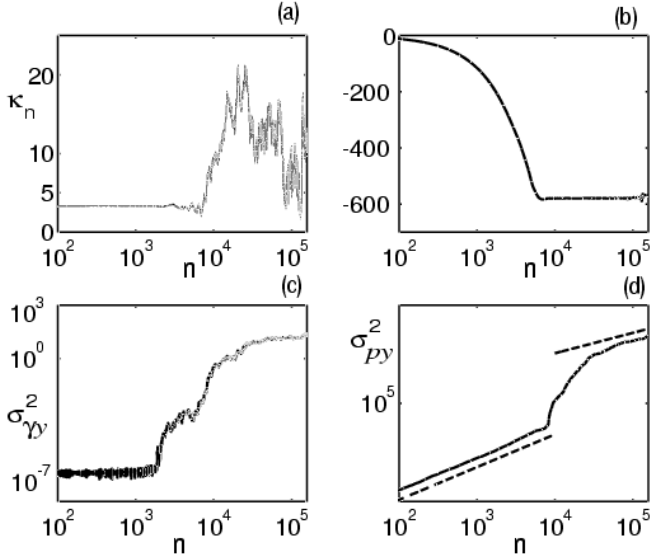,width=\columnwidth,angle=0}
\caption{Macroparticle instability and diffusion of passive scalar
concentration in the
fully self-consistent map
(\ref{D_map}) for a Gaussian distributed active scalar according to
Eq.~(\ref{gaussian}) and initial
conditions $\kappa(1)=3.3$ and $\theta(1)=0$. Panel (a) shows the
time series of $\kappa_n$, (b) the
phase $\theta_n$, (c) the square of the concentration variance
$\sigma^2_{\gamma y}$, and (d) the
square of the particle variance $\sigma^2_{p y}$. The lower and upper 
dashed lines have
slopes equal to $1$ and $0.62$ respectively.  Around
$n\approx 5
\times 10^3$, the
macroparticle   looses coherence, $\kappa_n$ exhibits large
fluctuations, $\theta_n$ drops to a
constant value,  and  there is a jump in
$\sigma^2_{\gamma y}$ and
$\sigma^2_{p y}$.}
\label{fig_macropart_insta}
\end{figure}
%%%%%%%%%%%%%%%%%%%%%%%%%%%%%%%%%%%%%%%%%%%%%%%%%%%%%%%%%%%%%%%%%%%%%%%%
%\vspace{2cm}
\begin{figure}
\epsfig{figure=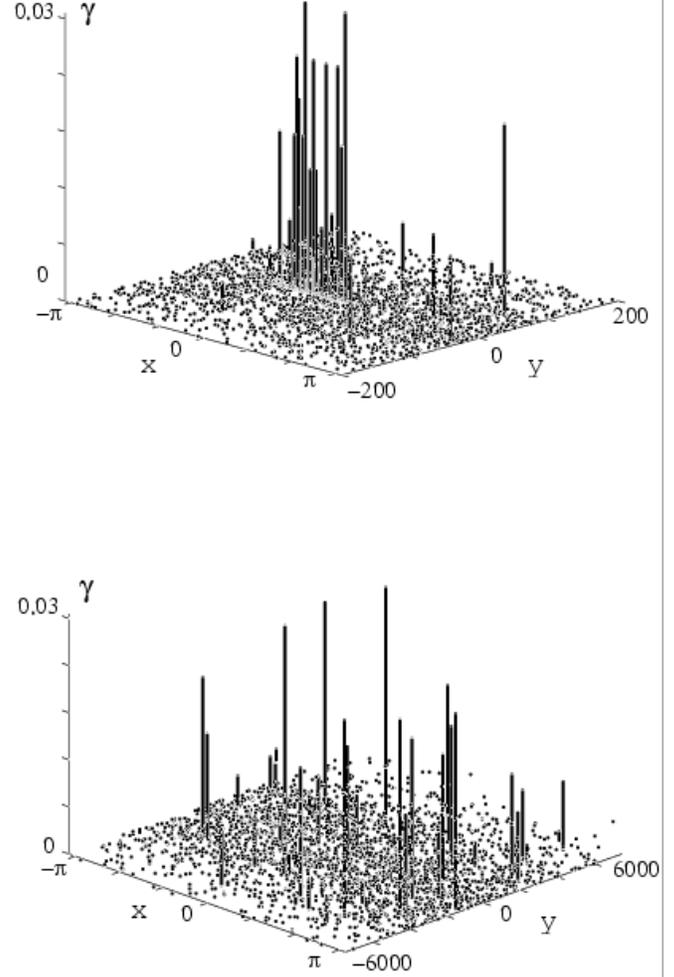,width=\columnwidth,angle=0}
\caption{Active scalar distribution during the
macroparticle instability phase (panel (a)), and the diffusive phase
(panel (b)) in the
fully self-consistent map (\ref{D_map}) for the run shown in
Fig.~\ref{fig_macropart_insta}.
The initial condition corresponds to the Gaussian distributed profile
in Eq.~(\ref{gaussian}) shown in
Fig.~\ref{fig_cluster_below}.  In the plots, the height of a vertical
line centered at $(x,y)$
corresponds to the active scalar concentration $\gamma_j$ of the
$j$-th particle located at
$(x,y)=(x_j,y_j)$.}
\label{fig_macropart_insta_phase}
\end{figure}
%%%%%%%%%%%%%%%%%%%%%%%%%%%%%%%%%%%%%%%%%%%%%%%%%%%%%%%%%%%%%%%%%%%%%%%%

\subsection{Quasilinear diffusion}

In the phase and amplitude coupled map there is also a
regime in which the phase is random
and the map is equivalent to a random standard map.
  This leads to quasilinear diffusion as shown in the following.
This regime
is approached for initial $\kappa(1)$ such that, in the long-time limit,
the time average value
$K_{eff}=1/(N_t-N_0)\sum_{n=N_0}^{N_{t}}\kappa(n)$ is larger than $1$,
being $N_t$ the final time, and $N_0$ a proper time chosen to avoid
the initial transient behavior.
Since the value of $\kappa $ depends on the iteration time,  the
eddy-diffusivity is also time dependent, and we must define an effective
diffusivity which is nothing but the time average of the
instantaneous diffusion,
$D_{eff}=1/(N_t-N_0)\sum_{n=N_0}^{N_t} D_y^E(n)$.
With these definitions,  the
quasi-linear approximation is recovered.
   This can be seen in Figure \ref{fig:QLDmap}
by plotting the $D_{eff}/D_{QL}$ vs $K_{eff}$, and with solid
line we plot the RPA approximation (\ref{eq:2.9}).
% For peculiar values of $K_{eff}$
%one has large $D$ which are probably the {\it ghosts} of the superdiffusion
%for the uncoupled map.
%%%%%%%%%%%%%%%%%%%%%%%%%%%%%%%%%%%%%%%%%%%%%%%%%%%%%%%%%%%%%%%%%%%%%%
%%%%%%%%%%
\vspace{2cm}
\begin{figure}
\epsfig{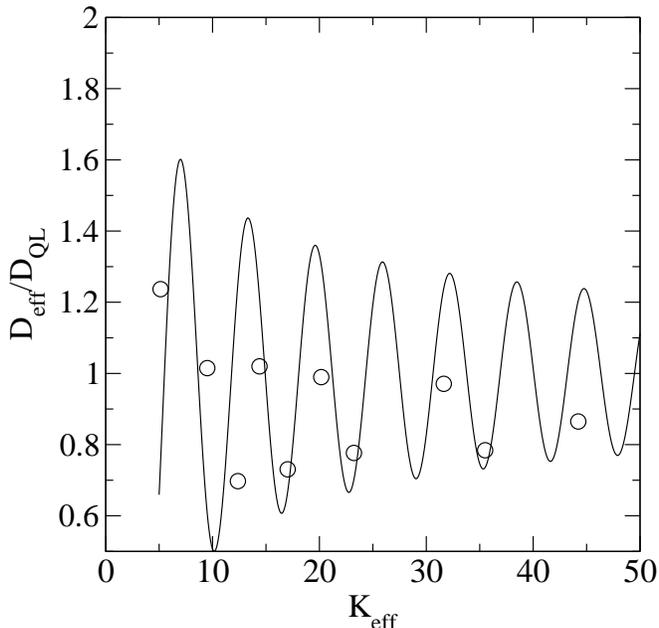}
\caption{With circles the numerically obtained values for
$D_{eff}/D_{QL}$ against $K_{eff}$. Solid line shows
the RPA approximation (\ref{eq:2.9}).
}
\label{fig:QLDmap}
\end{figure}
%%%%%%%%%%%%%%%%%%%%%%%%%%%%%%%%%%%%%%%%%%%%%%%%%%%%%%%%%%%%%%%%%%%%%%
%%%%%%%%%%

%%%%%%%%%%%%%%%%%%%%%%%%%%%%%%%%%%%%%%%%%%%%%%%%%%%%%%%%%%
\section{Conclusions}
\label{sec6}

We have studied self-consistent diffusion in active system
described by globally coupled  symplectic maps.
We focused our analysis in two
systems: an ensemble of phase coupled  maps,
and an ensemble of maps coupled through the
phase and the amplitude. The latter model is a symplectic 
discretization of the single
wave model and as such represents a simplified description of 
self-consistent transport
in plasmas and fluids. Numerical results obtained with this model indicate that
self-consistency plays a critical role in the diffusive properties of 
the system.
In particular, a) for small initial values of the standard map parameter,
$\kappa(1)=0.8$, coherent oscillations of the active scalar give rise to 
periodic oscilations
of  $\kappa$ above and below the threshold for barrier destruction 
($\kappa_c=0.9716$)
leading to subcritical diffusion; b) for smaller initial values, 
$\kappa(1)=0.6$, there is a
diffusive transient that eventually is suppressed by self-consistent 
effects; c) for larger
values, $\kappa(1)=3.3$, the active scalar loses coherence and this 
leads to a jump in the
particle diffusion and to the diffusion of the concentration field; 
d) for large enough
initial values of $\kappa$ there is widespread chaos and (with the 
exception of initial
values close to accelerator models)  self-consistency is shadowed by
stochasticity leading to  quasilinear difusion.  The behavior of the 
phase coupled map is in
general different. In this case, we have shown that a) in  the limit 
of large $K$, that is
strong stochasticity of the flow,  the external self-consistent field 
(at least for diffusion
properties) is equivalent to a random field; b) the singular 
properties of the  standard map,
like the existence of ballistic modes giving rise to anomalous diffusion, are
overcome by the external field; c) for $K$ small, diffusion is 
different from zero, i.e.
the external field breaks the barriers to transport present in the
standard map. Moreover, again the external field is equivalent to a 
random driving field.

\section{Acknowledgments}
We acknowledge discussions with Yves Elskens.
C.L. acknowledges support from the Spanish MECD.
D.dC.N. was supported by the Oak Ridge National Laboratory, managed by
UT-Battelle, LLC, for the U.S. Department of Energy under Contract
No.~DE-AC05-00OR22725.
D.dC.N. gratefully acknowledges the hospitality of the Department of
Physics of the University of Rome ``La Sapienza" during the elaboration
of this work. G.B. and A.V. had been supported by MIUR (Cofin.
Fisica Statistica di Sistemi Classici e Quantistici). A.V. acknowledges
support from INFM Center for Statistical Mechanics and Complexity.

%%%%%%%%%%%%%%%%%%%%%%%%%%%%%%%%%%%%%%%%%%%%%%%%%%%%%%%%%%

%\end{multicols}

\end{document}